\documentclass[twocolumn,showpacs,preprintnumbers,amsmath,amssymb]{revtex4}
\usepackage{graphicx}
\usepackage{dcolumn}

      \def\Bra#1{\left<#1\right|}
      \def\Ket#1{\left|#1\right>}
      {\catcode`\|=\active\gdef\Braket#1{\left<\mathcode`\|"8000\let|\bravert {#1}\right>}}
      \def\bravert{\egroup\,\vrule\,\bgroup}
      
\def\s{\sigma}

\begin{document}

\title{QUANTUM KEY DISTRIBUTION PROTOCOL WITH PRIVATE-PUBLIC KEY }
\author{Eduin Esteban Hernandez Serna}
\email[]{ceo@okphi.com}

\begin{abstract}
A quantum cryptographic protocol based in public key cryptography combinations and private key cryptography is presented. 
Unlike the BB84 protocol \cite{1} and its many variants\cite{2,3}, two quantum channels are used. The present research does not make reconciliation mechanisms of information to derive the key. A three related system of key distribution are described.
\end{abstract}
\maketitle

\section{Introduction}
In cryptography, the objective is to transmit information between two parties (Alice, Bob) that restrict access to an eavesdropper (Eve). In classical cryptography, the information is encrypted by a key which is kept in a secret or public way. The key is distributed among Alice and Bob to decrypt the message. Distribution the key remains a difficult issue. Quantum mechanics provides solutions with protocols that are largely determined by the following phases: preparation (Alice), measurement (Bob), verification and key derive (Alice and Bob). The last phase made public some details of the phases of preparation or measuring, this is called reconciliation mechanisms of information.
\\
This paper presents a quantum protocol based on public-private key cryptography for secure transmission of data over a public channel. The security of the protocol derives from the fact that Alice and Bob each use secret keys in the multiple exchange of the qubit. Unlike the BB84 protocol \cite{1} and its many variants\cite{2,3}, Bob knows the key to transmit, the qubits are transmitted in only one direction and classical information exchanged thereafter, the communication in the proposed protocol remains quantum in each stage. In the BB84 protocol, each transmitted qubit is in one of four different states, in the proposed protocol, the transmitted qubit can be in any arbitrary states.
\section{Protocol} 

\begin{itemize}
\item  	Alice took a bit $i$ transforming it in to an element of a secret base $B_{k}$ genering the qubit  $\Ket{\psi_{i,k}}$, that sends to Bob through a quantum channel .
\item 	Bob applies  $U_{j}$ that is only known by him, to the qubit $\Ket{\psi_{i,k}}$, returns the resulting qubit to Alice .
\item  Alice measures the qubit in the base $B_{k}$ obtaining the bit $j$ .
\end{itemize}

Where $B_{k}=\left\{\Ket{\psi_{0,k}},\Ket{\psi_{1,k}}\right\}$, 
\[
\begin{array}{ccc}
i=\Ket{\psi_{i,k}}& , & \Ket{\psi_{i,k}}=\alpha_{k}\Ket{i} +
(-)^{i}\beta_{k}\Ket{1-i}  \\
\end{array}
\]
\[
\begin{array}{ccc}
i \oplus(1-i)=1& , &\left|\alpha_{k}\right|^{2} +
\left|\beta_{k}\right|^{2} = 1 \\
\end{array}
\]

Initially, the qubit is in a pure state represented by the operator density $\rho=\Ket{\psi_{i,k}}\Bra{\psi_{i,k}}$ \cite{6} 
 when the qubit interacts whit the environment it becomes $$\rho^{'}=\sum_{j}E_{j}\rho E^{\dagger}_{j}$$

Where  $E_{j}$ operators acting in the space of a qubit. Therefore, $E_{j}=\s_{j0}I+\s_{j1}X+\s_{j2}Z+\s_{j3}XZ$ satisfy all conditions to $$ \sum_{j}E^{\dagger}_{j}\rho E_{j}=I $$ These operators transform the state of the qubit $\Ket{\psi_{i,k}} $in the overlap 
$$\Ket{\psi_{i,k}}\rightarrow E_{j}\Ket{\psi_{i,k}}$$
representing bit-flip, phase-flip, or both \cite{4,5}
\\
 bit-flip $X\Ket{\psi_{i,k}}$ 
$$
X\Ket{\psi_{i,k}}=\alpha_{k}\Ket{1-i} + (-)^{i}\beta_{k}\Ket{i}
$$ 

phase-flip $Z\Ket{\psi_{i,k}}$ 
$$
Z\Ket{\psi_{i,k}}=(-)^{i}\alpha_{k}\Ket{i}  - \beta_{k}\Ket{1-i}
$$

otherwise the error of both $XZ\Ket{\psi_{i,k}}$ or $ZX\Ket{\psi_{i,k}}$
\begin{center}
$XZ\Ket{\psi_{i,k}}=(-)^{i}\alpha_{k}\Ket{1-i}  - \beta_{k}\Ket{i}$
\\
$ ZX\Ket{\psi_{i,k}}=(-)^{1-i}\alpha_{k}\Ket{1-i} + \beta_{k}\Ket{i} $
\end{center}

give as measure  $1-i$ equivalent to the measured qubit $\Ket{\psi_{1-i,k}}$ in the base $B_{k}$. 
\\
Using this, Bob denotes operation  $U_{0}$  and  $U_{1}$ to the identity operator  $I$ and $XZ$ operator, respectively, where the bit $j$ lis responsible for operation $U_{j}$ 

The bit $j$, which was chosen by Bob and transmitted over a public channel, has reached Alice. Eve, the eavesdropper, cannot obtain any information by intercepting the transmitted qubits, although she could disrupt the exchange by replacing the transmitted qubits by her own. This can be
detected by:

\begin{itemize}
	\item 	appending parity bits, and/or
	\item   appending previously chosen bit sequences, which could be the destination and sending addresses or their hashed values, or some other mutually agreed sequence.
\end{itemize}

Since the $B_{k}$ and $U_{j}$ transformations can be changed as frequently as one pleases, Eve cannot obtain any statistical clues to their nature by intercepting the qubits.

%%%%%%%%%%%%%%%%%%%%%%%%%%%%%%%%%%%%%%%%%%%%%%%%%%%%%%%%%%%%%%%%%%%%%%%%%%%%%%%%%%%%%%%%%%%%%%%%%%%%%%%%%%%%%%%%%%%%%%%%%%%%%%%%
%%%%%%%%%%%%%%%%%%%%%%%%%%%%%%%%%%%%%%%%%%%%%%%%%%%%%%%%%%%%%%%%%%%%%%%%%%%%%%%%%%%%%%%%%%%%%%%%%%%%%%%%%%%%%%%%%%%%%%%%%%%%%%%%
%%%%%%%%%%%%%%%%%%%%%%%%%%%%%%%%%%%%%%%%%%%%%%%%%%%%%%%%%%%%%%%%%%%%%%%%%%%%%%%%%%%%%%%%%%%%%%%%%%%%%%%%%%%%%%%%%%%%%%%%%%%%%%%%
\section{Three Pseudo-code the protocol} 
The protocol is divided into three regions
\begin{description}
	\item I	\textit{	Preparation Phase (Alice)}
	\item II \textit{	Preparation Phase Key-Message (Bob)}
	\item III\textit{	Measurement and derivation phase (Alice) }
\end{description}

Eva is not involved in step I, II, because the \textit{key-message} has not been built, it only concern to be involved in step II, III. Three pseudo-codes are presented. The first is the ideal case where there are no errors, the second and third solve the quantum noise problem, gates errors and intervention from an eavesdropper (Eve).

The phase of preparation is the same for all pseudo-codes:
\\
\\
\textit{ I Phase}
\begin{itemize}

 	\item    Alice creates a randomly $tN$  bits, generating a string  $a$.
 	\item    Alice randomly chooses  $tN$ bases from the  $ \left\{B_{1}, B_{2},...,B_{n}\right\}$ generating a string  $b$.
 	\item    FOR each bit  $a_{k}$ in $a$ codify $a_{k}$  in the base $b_{k}$  resulting in the k-qubit.
 	\item    Alice send the generated qubits to Bob.
\end{itemize}
\ \\
 \textbf{Pseudo-code  1 }\\
\textit{ II Phase}
\begin{itemize}
	\item   Bob builds a binary string  $m$ \textit{Key-Message} $tN$ bits long. 
  	 \item    FOR each  $m_{k}$  in $m$ 
     \begin{itemize}
 		 	\item  IF ($m_{k} =1$) 		
 		   \begin{itemize}  		
 			  \item  Bob applies the gate to the $XZ$ or $ZX$ k-qubit. 	
 		   \end{itemize}			
 		  \end{itemize}
  	\item    Bob sends  $tN$ qubits to Alice.	
  	
\end{itemize}	
\textit{ III Phase}
\begin{itemize}
  	\item   Alice measures the k-qubit in the $B_{k}$ base generating a string $c$.
  	\item  Alice sum $c \oplus a$ obtaining the string $m$ \textit{Key-Message} used by Bob.
\end{itemize}
%%%%%%%%%%%%%%%%%%%%%%%%%%%%%%%%%
\ \\

\textbf{Pseudo-code  2 }\\
\textit{ II Phase}
\begin{itemize}

 		\item    Bob builds a binary string  $m$ \textit{Key-Message }$N$ bits long. 
 		\item    FOR each  $m_{k}$  in $m$ 
      \begin{itemize}
 			 \item	 IF ($ m_{k}=1$)
 			 	\begin{itemize}
 			 	 \item  Bob applies the gate to the $XZ$ or $ZX$ from the  $\left(kt +1-t\right)$-qubit to $kt$-qubit.
 			 	\end{itemize}
 			\end{itemize}	
 		\item   Bob sends  $tN$ qubits to Alice.

\end{itemize}	
\textit{ III Phase}
\begin{itemize}

 		\item    Alice measures the $k$-qubit in the   $b_{k}$ base generating a string $c$.
 		\item    Alice sum $c \oplus a$ obtaining the string  $M $ and build the string $d_s = M_{st +1 - t } M_{st+2 - t},…, M_{st}$   with  $s=\left\{1,2,...,N\right\}$ where the string $M=d_{1}, d_{2},…, d_{ N } $.
 		\item   FOR each string  $d_{ s }$ of $M$ will be generated a string $m^{'}$ and $p$, $N$ bits long as follows.
	    \begin{itemize}
			 \item  IF ($f (d_ {s}, i)>t$), where $f (d_ {s}, i)$ counts the bit $i$ at the $d_{ s }$ string
				 \begin{itemize} 
					 \item  $m^{'}_ {s}=i$ y $p_ {s}=0$ 
				 \end{itemize} 
	  	\end{itemize} 
			\begin{itemize} 					 								
				\item	ELSE
				 \begin{itemize} 
					\item  $p_{s}=1$ y $m^{'}_ {s}=0$ 
				 \end{itemize} 
			\end{itemize} 
	 \item    Alice send to Bob the string  $p$.
	 \item    FOR each $p_{s}$ and some  $p_{k}=0$ in $p$ 
		 \begin{itemize}
			 \item IF ($p_{s}=1$ )
				 \begin{itemize} 
				  \item Alice  $m^{'}_{k}\oplus m^{'}_{s}$ of $m^{'}$ generating the string $C$ \textit{key-message}
					\item Bob    $m_{k}\oplus m_{s}$ of $m$ generating the string $C$ \textit{key-message}
				 \end{itemize} 	
			\end{itemize}
\end{itemize}
%%%%%%%%%%%%%%%%%%%%%%%%%%%%%%%%%%%%%%%%%%%%5
\ \\

\textbf{Pseudo-code  3 }\\
\textit{ II Phase}
\begin{itemize}
 		\item   Bob builds a binary string  $m$ \textit{Key-Message} $N$ bits long. 
 		\item   FOR each N-qubits 
 		\begin{itemize} 
     \item FOR each  $m_{k}$  in $m$ 
	    \begin{itemize} 
			 \item IF ($m_{k} =1$)
				\begin{itemize} 
 				 \item Bob applies the gate to the $XZ$ or $ZX$ k-qubit.
 				\end{itemize}
 			\end{itemize}
 	  \end{itemize}	   					 		
 		\item   Bob sends  $tN$ qubits to Alice.

\end{itemize}	
\textit{ III Phase}
\begin{itemize}

  \item   Alice measures the $k$-qubit in the   $b_{k}$ base generating a string $c$.
  \item    Alice sum $c \oplus a$ obtaining the string  $M $ and build the string $d_s = M_{st +1 - t } M_{st+2 - t},…, M_{st}$   with  $s=\left\{1,2,...,N\right\}$ where the string $M=d_{1}, d_{2},…, d_{ N } $.
  \item    Alice compares all the binary strings $d_ {s}$ and rebuilds  $m$.
\end{itemize}
	
%%
%%%%%%%%%%%%%%%%%%%%%%%%%%%%%%%%%%%%%%%%%%%%%%%%%%%%%%%%%%%%%%%%%%%%%%%%%%%%%%%%%%%%%%%%%%%%%%%%%%%%%%%%%%%%%%%%%%%%%%%%%%%%%%%%
%%%%%%%%%%%%%%%%%%%%%%%%%%%%%%%%%%%%%%%%%%%%%%%%%%%%%%%%%%%%%%%%%%%%%%%%%%%%%%%%%%%%%%%%%%%%%%%%%%%%%%%%%%%%%%%%%%%%%%%%%%%%%%%%
%%%%%%%%%%%%%%%%%%%%%%%%%%%%%%%%%%%%%%%%%%%%%%%%%%%%%%%%%%%%%%%%%%%%%%%%%%%%%%%%%%%%%%%%%%%%%%%%%%%%%%%%%%%%%%%%%%%%%%%%%%%%%%%%
\section{GENERALIZATION} 
Only Bob is involved in the preparation phase of \textit{Key-Message} this allows extended to three parties (Alice, Bob, Celine) unlike the standard protocols \cite{7}.

\begin{description}	
	\item I	\textit{	Preparation Phase (Alice,Celine)}
	\item II \textit{	Preparation Phase Key-Message (Bob)}
	\item III\textit{	Measurement and derivation phase (Alice, Celine) }
\end{description}

\begin{itemize}
	\item  	Alice took a bit  $i$ transforming it in to an element of a secret base $B_{k}$,  Celine took a bit  $i$ transforming it in to an element of a secret base $B_{t}$,  both sends their qubit to Bob through a quantum channel.
	\item  	Bob applies  $U_{j}$ secret operation on the qubits  $\Ket{\psi_{i,k}}$ and $\Ket{\psi_{s,t}}$ returns qubits 
resulting to their respective parties
	\item  Alice and Celine measures the qubits in the base $B_{k}$ and $B_{t}$  obtaining a value sent by Bob. 
\end{itemize}

Generalizing the protocol for $n$ parties, where Bob is central, a quantum key-message distribution network will be obtained.

\section{CONCLUSIONS } 
The quantum protocol presented with its variants provides a safe sending of information of direct communication between two or more parties. The generalizations for n parties can create a network of massive sending information for $n -1$ parties being one of them the key-message distribution center. This protocol is used to distribute applications key-messages safe over long distances because it allows the sending of massive qubits.
Since the proposed protocol does not use classical communication, it is immune to the man-in-the-middle attack on the classical communication channel which BB84 type quantum cryptography protocols suffers from \cite{8}. On the other hand, implementation of this protocol may be harder because the qubits get exchanged multiple times.

\end{document}